\documentclass[preprint,numbers,compress]{elsarticle}
\usepackage{amssymb}
\usepackage{amsmath}
\usepackage{float}
\usepackage{pdflscape}
\usepackage{multirow}

\journal{Chemical Engineering Journal}
\begin{document}

\begin{frontmatter}
\title{Mechanistic Modeling of Lipid Nanoparticle (LNP) Precipitation via Population Balance Equations (PBEs)}

\author[inst1]{Sunkyu Shin}
\author[inst1]{Cedric Devos}
\author[inst1]{Aniket Pradip Udepurkar}
\author[inst1]{Pavan K. Inguva}
\author[inst1]{Allan S. Myerson}
\author[inst1]{Richard~D.~Braatz\corref{cor1}}

\cortext[cor1]{Corresponding author. Email: \texttt{braatz@mit.edu}}

\address[inst1]{Department of Chemical Engineering, Massachusetts Institute of Technology, 77 Massachusetts Avenue, Cambridge, MA 02139, United States}

\begin{abstract}
Lipid nanoparticles (LNPs) are precisely engineered drug delivery carriers commonly produced through
controlled mixing processes, such as nanoprecipitation. Since their delivery efficacy greatly depends on
particle size, numerous studies have proposed experimental and theoretical approaches for tuning LNP size.
However, the mechanistic model for LNP fabrication has rarely been established alongside experiments, limiting a profound understanding of the kinetic processes governing LNP self-assembly. Thus, we present a population balance equation (PBE)-based model that captures the evolution of the particle size distribution (PSD) during LNP fabrication, to provide mechanistic insight into how kinetic processes control LNP size. The model showed strong agreement with experimentally observed
trends in the PSD. In addition to identifying the role of each kinetic process in shaping the PSD, we analyzed
the underlying mechanisms of three key operational strategies: manipulation of (1) lipid concentration, (2)
flow rate ratio (FRR), and (3) mixing rate. We identified that the key to producing precisely controlled
particle size lies in controlling supersaturation and lipid dilution to regulate the balance between nucleation
and growth. Our findings provide mechanistic understanding that is essential in further developing strategies
for tuning LNP size.
\end{abstract}

\begin{highlights}

\item A mechanistic model for lipid nanoparticle (LNP) fabrication is developed based on population balance equations (PBEs).
\item The model successfully reproduces the experimentally measured particle size distribution (PSD) of LNPs.
\item Three key operational strategies for controlling particle size are analyzed in terms of kinetic processes.

\end{highlights}


\end{frontmatter}

\section{Introduction}

Lipid nanoparticle (LNP) technology has emerged as a promising drug delivery system after it demonstrated its effectiveness during the COVID-19 pandemic. One advantage of the LNP system is that particles can be precisely tuned by modifying its structure and surface properties, allowing specific organ-targeted delivery \cite{houLipidNanoparticlesMRNA2021, wittenArtificialIntelligenceguidedDesign2024, huangOrganselectiveLipidNanoparticles2024}.  
One of the most critical quality attribute of LNPs is the particle size distribution (PSD), as it directly affects biodistribution within the body \cite{zhigaltsevBottomUpDesignSynthesis2012}. Particles smaller than 100--$200~\mathrm{nm}$ are generally preferred, as they exhibit enhanced permeation through biological barriers and improved accumulation in target tissues, mainly due to their efficient drainage to lymph nodes \cite{kimuraDevelopmentMicrofluidicBasedPostTreatment2020, zhigaltsevBottomUpDesignSynthesis2012, okudaSizeregulationRNAloadedLipid2022}. Meanwhile, larger particles are preferred when targeting liver sinusoidal endothelial cells (LSECs) \cite{satoRelationshipPhysicochemicalProperties2016}. Additionally, larger particles have advantages in targeting antigen-presenting cells (APCs), as they are more efficiently taken up by these cells, particularly through macropinocytosis in dendritic cells \cite{jiLipidMicroparticlesShow2023, okudaSizeregulationRNAloadedLipid2022}. Larger particles were reported to elicit a stronger T cell response \cite{watsonDesignConsiderationsLiposomal2012}. These findings highlight the need for tuning LNP size, with consideration of the target organ and therapeutic effects \cite{hassettImpactLipidNanoparticle2021, maekiUnderstandingFormationMechanism2017}. Therefore, a manufacturing system capable of producing LNPs with a tunable size is crucial to meet the broad range of LNP applications for therapeutic uses.

The main challenges in tuning the size of LNPs arise from their complex nature, in which multiple physicochemical phenomena are interconnected. For example, LNPs are typically fabricated through the self-assembly of four major lipid components: ionizable lipids, polyethylene glycol (PEG)-lipids, phospholipids, and cholesterol  \cite{ripollOptimalSelfassemblyLipid2022}. This process is driven by the rapid mixing of an organic phase with an aqueous buffer, which rapidly increases the polarity of the organic phase. This sudden polarity shift induces lipid self-assembly into a dense core surrounded by lipid bilayers, encapsulating the payload \cite{leungLipidNanoparticlesContaining2012, leungMicrofluidicMixingGeneral2015}. The formation of LNPs is governed by the instantaneous solvent environment, characterized by inhomogeneous mixing, posing significant challenges to precise engineering.

Recently, microfluidic systems have been introduced to enable precise control over LNP size through refined manipulation of mixing patterns at the micrometer scale. Compared to bulk mixing systems, microfluidic systems offer superior scalability and reproducibility due to their high precision and ability to operate in parallel \cite{pratsinisImpactNonionizableLipids2023, ripollOptimalSelfassemblyLipid2022,zookEffectsTemperatureAcyl2010, shepherdScalableMRNASiRNA2021}. In addition, this technology has enabled mechanistic insights into several key operational strategies. For example, Okuda et al.\ \cite{okudaSizeregulationRNAloadedLipid2022} experimentally demonstrated that salt concentration is a critical factor in producing small-sized LNPs. Maeki et al.\ \cite{maekiMicrofluidicTechnologiesDevices2022} and Hamdallah et al.\ \cite{hamdallahMicrofluidicsPharmaceuticalNanoparticle2020} reported that a higher flow rate ratio (FRR)—defined as the ratio of aqueous to organic phase—significantly reduces LNP size, by producing faster dilution of lipid, thereby suppressing the growth of lipid into larger particles. Kimura et al.\ \cite{kimuraDevelopmentMicrofluidicBasedPostTreatment2020} and Pradhan et al.\ \cite{Pradhan2008123123} reported that particle size decreased with increasing total flow rate (TFR). More systematic methods, such as the design of experiments \cite{teradaCharacterizationLipidNanoparticles2021, maharjanComparativeStudyLipid2023} and regression modeling \cite{okudaSizeregulationRNAloadedLipid2022}, have also been proposed to quantitatively measure the impact of each operational strategy.

While experimental studies have provided valuable insights into LNP structures and the influence of operational parameters on their size, these findings have rarely been incorporated into mechanistic modeling frameworks.
Consequently, the kinetic principles underlying various operational strategies are seldom addressed in modeling studies. Furthermore, economic viability and scalability are critical for the successful medical deployment of LNPs \cite{balbinoContinuousFlowProduction2013}. This highlights the need for mechanistic model as a tool to strictly optimize LNP manufacturing.
LNP fabrication, in which lipids self-assemble triggered by a polarity shift, is analogous to antisolvent crystallization and is often referred to as nanoprecipitation
\cite{martinezrivasNanoprecipitationProcessEncapsulation2017}. 
Therefore, population balance equations (PBEs), which have been used effectively to predict the evolution of PSD in various crystallization processes \cite{raponiDeepLearningKinetics2024, nagyModelingPharmaceuticalFiltration2021, hanhounSimultaneousDeterminationNucleation2013}, may provide mechanistic framework for modeling the formation of LNPs. Despite their versatility, PBEs have rarely been applied to LNP fabrication \cite{inguvaMechanisticModelingLipid2024}.

To this end, we present a mechanistic model to investigate how operational strategies affect LNP size control. Specifically, we formulate PBEs based on nucleation, growth, and coalescence kinetics, which are commonly recognized as key mechanisms in crystallization. The model is used to reproduce the evolution of PSD measured experimentally over time, showing good agreement. Then the model is used to systematically elucidate the role of each kinetic process in PSD shape. In addition, the effects of key operational strategies—including lipid concentration, FRR, and mixing rate—are analyzed by investigating their underlying kinetic mechanisms. Based on key principles to enable precise particle size tuning, we evaluate the advantages of each operational strategy. We believe that this study provides a mechanistic understanding of LNP size control, thereby facilitating efficient and flexible LNP manufacturing. 

\section{Methods}

\subsection{Experimental setup}
\begin{figure}[H]
    \centering
    \includegraphics[width=0.9\textwidth]{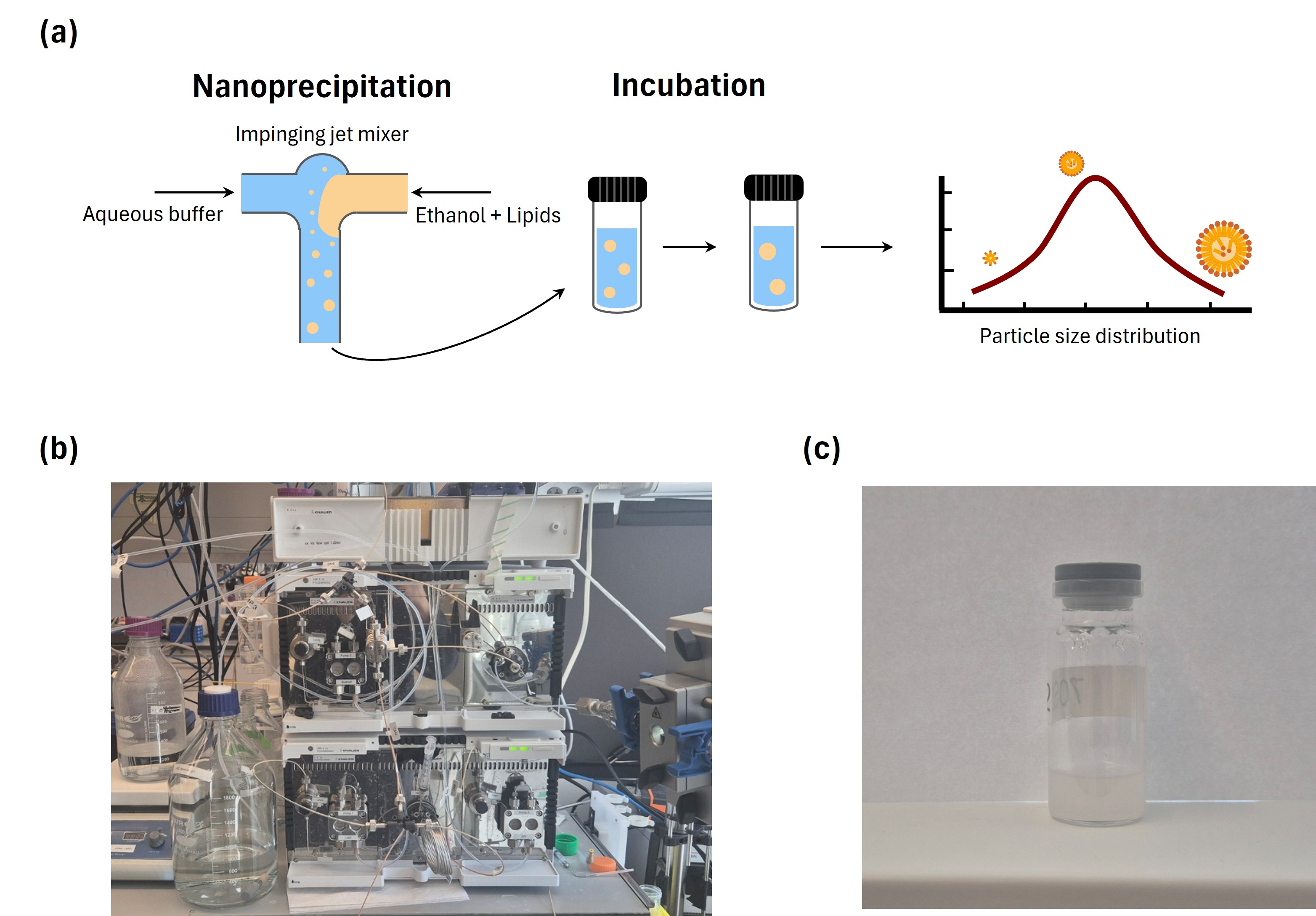}
    \caption{Schematic diagram of (a) the LNP fabrication process, (b) the microfluidic IJM system, and (c) the resulting lipid nanoparticles (LNPs).}
    \label{fig:Fig. 1}
\end{figure}

We experimentally synthesized LNPs via nanoprecipitation using a microfluidic impinging jet mixer (IJM) system. The schematic diagram of the experimental setup is shown in Fig.~\ref{fig:Fig. 1}a. The lipid phase consisted of four components: ionizable lipid ($50~\mathrm{mol}\%$), cholesterol ($38.5 ~\mathrm{mol}\%$), phospholipid ($10~\mathrm{mol} \%$), and PEG lipid ($1.5 ~\mathrm{mol}\%$). The ionizable lipid used was 1,2-dioleyloxy-3-dimethylaminopropane (DODMA). These lipids were dissolved in ethanol to achieve a lipid concentration of $10~\mathrm{mg}~\mathrm{mL}^{-1}$ as the default case.
As an antisolvent, an aqueous buffer solution containing 0.1 M sodium acetate at pH 5.5 was prepared. LNPs were fabricated by introducing both streams at a flow rate of $4~\mathrm{mL}~\mathrm{min}^{-1}$ under FRR of 3 in the IJM NanoScaler (KNAUER Wissenschaftliche Geräte GmbH, Berlin, Germany). The IJM system and fabricated LNPs are shown in Figs.~\ref{fig:Fig. 1}bc. The crude LNP product from the IJM was incubated in a temperature-controlled chamber, and its PSD was measured over time using dynamic light scattering (DLS) with the NanoFlowSizer (InProcess-LSP, The Netherlands).

\subsection{Modeling strategy}

\begin{figure}[H]
    \centering
    \includegraphics[width=0.9\textwidth]{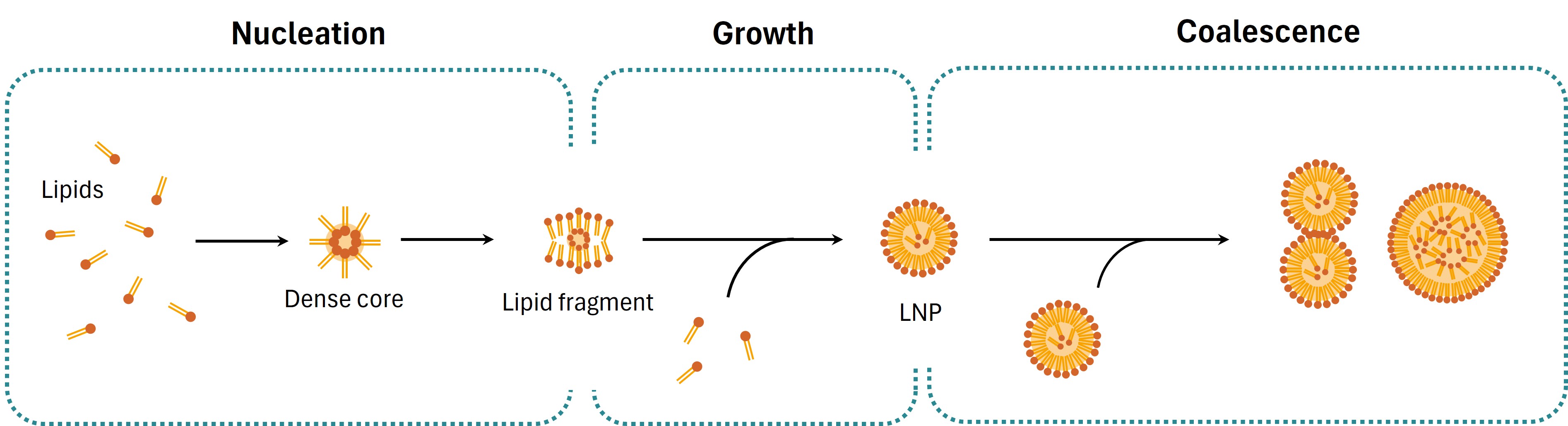}
    \caption{Schematic diagram of LNP formation mechanism. }
    \label{fig:Fig. 2}
\end{figure}

We developed a residence time-based model to consistently interpret all steps of the experimental procedure. In the IJM, the two streams are rapidly mixed at the junction point and subsequently flow through the downstream channel, which can be interpreted as a tubular reactor. The subsequent incubation step following mixing can be considered a batch reactor. Since models for both reactors are solved identically in terms of residence time, a residence time-based calculation enables seamless incorporation of these two steps.

Rapid precipitation processes at the nanoscale have been understood in terms of crystallization kinetic processes due to their thermodynamic similarities in driving forces and kinetics \cite{martinezrivasNanoprecipitationProcessEncapsulation2017, myersonHandbookIndustrialCrystallization2002, hamdallahMicrofluidicsPharmaceuticalNanoparticle2020}.
The detailed mechanisms of LNP fabrication are broken down and interpreted as crystallization kinetic processes in Fig.~\ref{fig:Fig. 2}.
In the initial stage, the lipid components assemble  in response to a sudden increase in supersaturation, forming a dense core with an inverted micelle structure \cite{kulkarniFusiondependentFormationLipid2019}. Almost immediately after, nearby lipid molecules further assemble to form lipid fragments with a bilayer structure. These two steps occur almost instantaneously and are analogous to nucleation events in crystallization \cite{maekiUnderstandingFormationMechanism2017}.
Lipid fragments are semi-stable and gradually grow by the continued integration of surrounding lipid molecules. This gradual process is governed by the supersaturation of lipids, which is the same driving force as growth kinetics in crystallization \cite{zookEffectsTemperatureAcyl2010, nguyenControlledNucleationLipid2012, martinezrivasNanoprecipitationProcessEncapsulation2017}.
Additionally, LNPs undergo spontaneous fusion or agglomeration upon binary collisions, leading to the formation of larger particles or clusters. This process can be described by coalescence kinetics using a collision kernel \cite{yangExperimentalInvestigationPopulation2012}.
In the subsequent sections, thus, the model for LNP fabrication will be implemented and analyzed in terms of these three crystallization kinetics processes.

\subsubsection{PBEs}

To estimate the temporal evolution of PSD during LNP fabrication, the PBE is formulated as
\begin{equation}
    \frac{\partial n(L)}{\partial t} = B_n + B_c + D_c - \frac{\partial (G n)}{\partial L}\text{,}
\end{equation}
where $n$ is the particle number density function, $B_n$ denotes the birth rate due to nucleation, $B_c$ and $D_c$ 
denote birth and death rate due to coalescence respectively, and $G$ denotes the kinetic growth rate. The raw measurement of PSD in DLS experiment yields a light intensity-weighted distribution where the number density function is weighted by the particle size to the power of 6. To enable comparison with the model results, the number density distributions from the model are converted to the intensity-weighted distribution, $n_\text{DLS}$, by
\begin{equation}
    n_\text{DLS}(L) = n(L) L^6\text{.}
\end{equation}
For numerical integration, the PBE is discretized along the particle size dimension using 200 logarithmically spaced bins. 
The number of bins was determined considering both computational cost and numerical accuracy. 

\subsubsection{Solubility model}

The primary driving force in antisolvent crystallization is the supersaturation of the solute, which in this study is lipid. Although LNPs are composed of four lipid types, the solubility data for each lipid component are scarcely reported in the literature, except for cholesterol, which accounts for $38.5~\mathrm{mol}\%$  in LNPs. To the best of authors’ knowledge, assuming the solubility of cholesterol for other lipid components offers a practical basis for modeling.

\begin{figure}[H]
    \centering
    \includegraphics[width=0.9\textwidth]{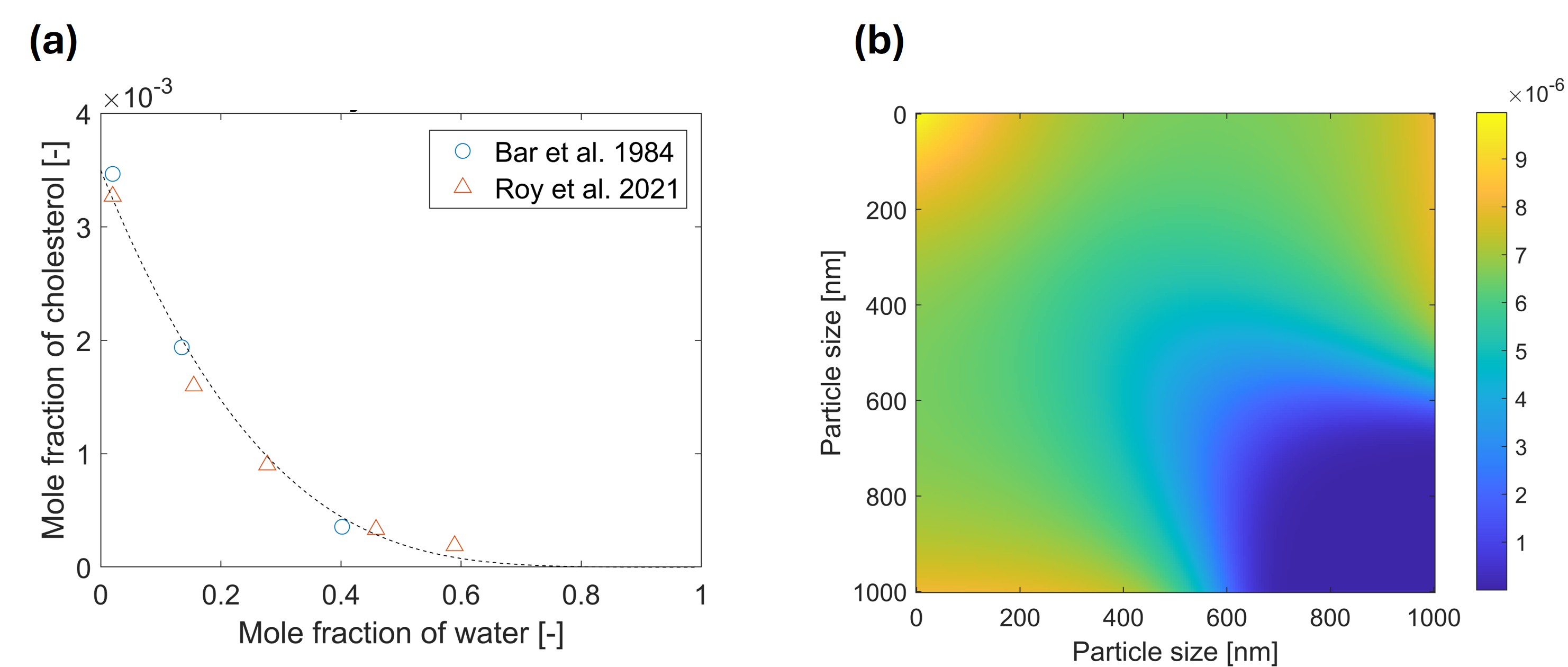}
    \caption{(a) Comparison of solubility model with experimental data from literature \cite{royAnalysisMechanismCholesterol2021} and (b) attachment efficiency calculated from DLVO theory.}
    \label{fig:Fig. 3}
\end{figure}

The solubility of cholesterol in a water-ethanol mixture is estimated using the extended version of Yalkowsky’s log-linear relationship \cite{mirheydariComparisonModelsCorrelation2019, millardSolubilizationCosolventsEstablishing2002}, 
\begin{equation}
\ln x = w_1 \ln x_1 + w_2 \ln x_2\text{,}
\end{equation}
as it exhibits satisfactory consistency with experimentally reported solubility data across various mixture ratios (Fig.~\ref{fig:Fig. 3}a), where $x$, $x_1$, and $x_2$ denote mole fraction solubility in the mixture, water, and ethanol, respectively, and $w_1$ and $w_2$ denote volume fraction of each solvent.

\subsubsection{Nucleation}
The nucleation model is developed to describe the rapid formation of the lipid fragments, which is analogous to nuclei formation \cite{maekiUnderstandingFormationMechanism2017}. 
The time scale of the formation of the individual lipid core and subsequent fragment ranges from microseconds to milliseconds at the molecular level \cite{leungLipidNanoparticlesContaining2012}.
The overall nucleation rate remains a rate-limiting kinetic process competing with growth, because the overall nucleation is frequently delayed by surrounding factors, such as molecular diffusion and local fluctuations arising from inhomogeneous mixing \cite{royAnalysisMechanismCholesterol2021, martinezrivasNanoprecipitationProcessEncapsulation2017}.
To reflect the stochastic nature of nucleation, a distributed nucleation size is frequently used instead of a fixed critical length \cite{inguvaMechanisticModelingLipid2024}. Specifically, this study employs the Gaussian kernel ($\mathcal{G}$) to distribute the overall nucleation rate ($B_0$) around the critical nucleation length ($L_c$), 
\begin{equation}
    B_n = B_0 \mathcal{G}(L, L_c)\text{,}
\end{equation}
\begin{equation}
    \mathcal{G}(L, L_c)  = \exp\left(-\frac{(L - L_c)^2}{2\sigma^2}\right)\text{,}
\end{equation} 
which allows the instantaneous emergence of PSD at the beginning of the simulation.
For the nucleation kinetic rate, classical nucleation theory (CNT), which assumes primary and homogeneous nucleation is employed, as it provides a thermodynamic explanation for determining the critical nucleation size ($L_C$) and the kinetic rate ($B_0$) \cite{myersonHandbookIndustrialCrystallization2002, choiPreciseControlLiposome2023}, 
\begin{equation}
    L_C = \frac{4\sigma V_m}{k_\mathrm{B} T \ln S}\text{,}
\end{equation}
\begin{equation}
    B_0 = A_n \exp\!\left(\!-\frac{16\pi \sigma^3 V_m^2}{3(k_\mathrm{B} T)^3 (\ln S)^2} \!\right)\!\text{,}
\end{equation}
where $\sigma$, $V_m$, $k_\mathrm{B}$, and $A_n$ denote interfacial energy, molecular volume, the Boltzmann constant, and the pre-exponential factor, respectively. While the CNT model is used here to describe nucleation, it is worth noting that various mechanisms may contribute to the initial formation of lipid fragments.
At extremely high supersaturation, spinodal decomposition can occur, where an unstable mixture is instantaneously decomposed into two phases. However, as supersaturation declines, the driving mechanism for solid phase formation transitions from spinodal decomposition to homogeneous nucleation. Indeed, homogeneous nucleation is the main driving mechanism in many nano-precipitation systems \cite{hamdallahMicrofluidicsPharmaceuticalNanoparticle2020}. 
Therefore, spinodal decomposition is not explicitly described in this study. 
Similarly, the minor contributions of secondary and heterogeneous nucleation are implicitly described using the CNT model with fitted parameters ($A_n$ and $\sigma$), due to their structural similarity to homogeneous nucleation.

\subsubsection{Growth}

A growth model is formulated to describe the successive enlargement of lipid fragments driven by supersaturation. Growth is generally limited by two kinetic steps: molecular diffusion of the solute and its subsequent integration at the crystal surface \cite{myersonHandbookIndustrialCrystallization2002, hamdallahMicrofluidicsPharmaceuticalNanoparticle2020}. 
To describe the complex growth mechanism, empirical models are often used due to their flexibility
\cite{royAnalysisMechanismCholesterol2021, zhangSolvingCrystallizationPrecipitation2024, galbraithModellingSimulationInorganic2014}. In this study, the Bransom model is used, which has a length term in the kinetic rate law \cite{zhangSolvingCrystallizationPrecipitation2024}. This mathematical form automatically allows a log-normal distribution due to the law of proportionate effect, without the need to assume a dispersed growth rate model \cite{eberlGeologicalInterpretationsCrystal2002}. To consider size-dependent solubility and the Ostwald ripening effect, the driving force term was modified to be driven by a difference between supersaturation ($S$) and equilibrium supersaturation ($S^*$), as suggested by Iggland and Mazzotti \cite{igglandPopulationBalanceModeling2012}, 
\begin{equation}
    G(L) = k_g L C_\text{lipid} (S - S^*)\text{,}
\end{equation}
where $k_g$ and $L$ represent kinetic parameter and particle size, respectively.

\subsubsection{Coalescence}
The coalescence model aims to describe the integration of two particles to form a larger structure. For this inter-particle interaction, various behaviors have been reported with inconsistent terms such as membrane fusion, aggregation, coagulation, and agglomeration \cite{yangExperimentalInvestigationPopulation2012, kulkarniFusiondependentFormationLipid2019, choiPreciseControlLiposome2023}.
Based on their similarity,
this study employs a coalescence mechanism to approximate various binary integrations as a lumped framework. 
Accordingly, the term agglomeration is used to refer to the physical behavior of LNPs, while coalescence refers to the kinetic process as represented in the numerical model.
The integration event is assumed to subsequently occur after the collision of two particles, driven by spontaneous Brownian motion \cite{schwarzerPredictionAggregationKinetics2005, yangExperimentalInvestigationPopulation2012}. The birth ($B_c$) and death terms ($D_c$) due to coalescence are given by 
\begin{equation}
B_c (L) = \frac{L^2}{2} \int_{0}^{L}  \frac{\beta \left(\sqrt[3]{L^3 - \lambda^3}, \lambda^3 \right) n\left(\sqrt[3]{L^3 - \lambda^3} \right) n(\lambda)}{(L^3 - \lambda^3)^{2/3}} d\lambda\text{,}
\end{equation}
\begin{equation}
D_c (L) = n(L) \int_{0}^{\infty} \beta(L, \lambda) n(\lambda) d\lambda\text{,}
\end{equation}
where $L$ and $\lambda$ are sizes of particles colliding, and $\beta$ is a kernel to describe kinetic rate of coalescence. The Smoluchowski collision kernel ($\beta_\text{Smoluchowski}$) is a popular model to describe collision frequency due to Brownian motion and has proven its efficacy in a range of nanoparticle systems \cite{traoreMechanisticInsightsSilvergold2024, casadoPredictingSizeSilver2024}. This kernel is a function of two particles ($L$ and $\lambda$), viscosity ($\mu$), and temperature ($T$), 
\begin{equation}
\beta_\text{Smoluchowski}(L, \lambda) = \frac{2 k_\mathrm{B} T (L + \lambda)^2}{3 \mu L \lambda}
\end{equation}

In addition to Brownian collisions, the impact of interparticle forces is considered by introducing attachment efficiency ($\alpha$) \cite{capcoNanomaterialImpactsCell2014}, 
\begin{equation}
\beta = \alpha\beta_\text{Smoluchowski}\text{.}
\end{equation}

Based on the Erjaguin–Landau–Verwey–Overbeek (DLVO) theory, which accounts for van der Waals attraction and electrostatic repulsion, the attachment efficiency is calculated as
\begin{equation}
    \alpha =  k_\text{PEG}\frac{(L+\lambda)}{2} \int_{0}^{\infty} \frac{\exp\!\left(\frac{\phi_{\text{total}}}{k_\text{B} T}\right)}{a^2} \text{d}a\text{,}
\end{equation}
\begin{equation}
    \phi_{\text{total}} = \phi_{\text{VDW}} + \phi_{\text{elec}}\text{,}
\end{equation}
where  $\phi_\text{total}$, $\phi_\text{VDW}$, and $\phi_\text{elec}$ denote total, van der Waals, and electrostatic energy barriers, respectively. We introduced a fitted parameter ($k_\text{PEG}$) to account for the effect of PEG-lipids, which are frequently used to prevent agglomeration of LNPs because their long tails induce steric repulsion between particles \cite{capcoNanomaterialImpactsCell2014}. 
The variable $a$ denotes inter-particle distance. Detailed implementation of each energy barrier is provided in the Supplementary Materials. Calculated attachment efficiency is presented in Fig.~\ref{fig:Fig. 3}b, indicating that smaller particles have higher efficiency value, which is because repulsive electrostatic effect can be more easily screened by the surrounding ions in smaller particles \cite{capcoNanomaterialImpactsCell2014}.

\section{Results and Discussion}

\subsection{Model validation and parameter estimation}

\begin{figure}[H]
    \centering
    \includegraphics[width=0.9\textwidth]{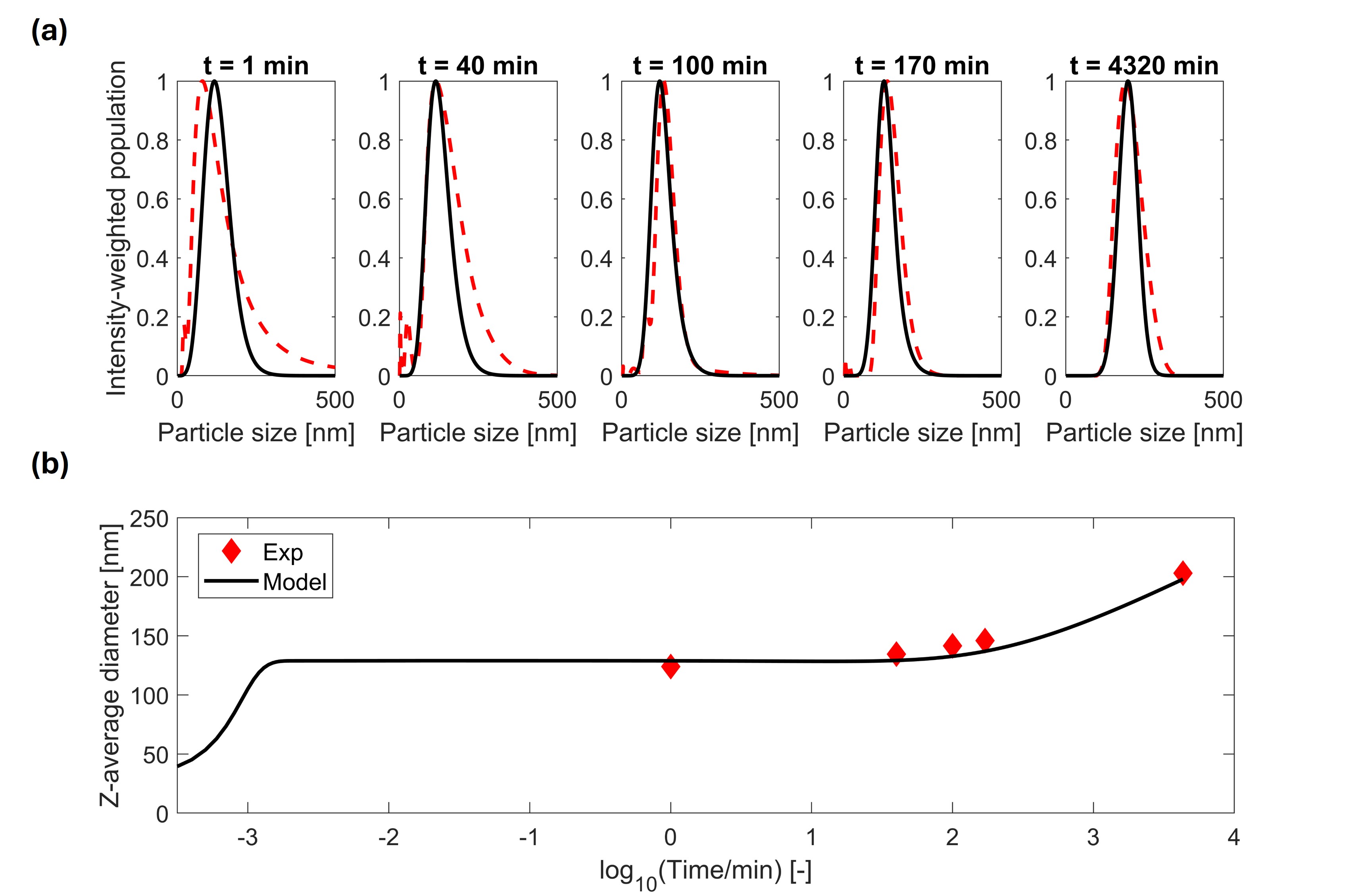}
    \caption{Comparison of model predictions and experimental measurements. (a) PSD with model prediction (black solid line) and experimental measurement (red dashed line). (b) Z-average diameter.}
    \label{fig:Fig. 4}
\end{figure}
The PSD was first measured shortly after mixing in the IJM, in about 1 min, followed by further incubation with additional measurement over time (for 4,320 min).
Kinetic parameters in the model were estimated by comparison to experimental DLS measurements. The PSD at 1 min reflects the result of rapid self-assembly regime in which nucleation and growth compete in the IJM. 
Since the relative balance between growth and nucleation is the key factor determining PSD and z-average diameter at this early stage, their relative magnitude can be estimated from the PSD at 1 min. 
Therefore, the ratio of nucleation ($A_n$) and growth ($k_g$) parameters was adjusted to ensure the simulated PSD at $1~\mathrm{min}$ matched the overall curve of measured PSD at $1~\mathrm{min}$.
On the other hand, the absolute scale of kinetic parameters ($A_n$ and $k_g$) of nucleation and growth was roughly estimated based on the assumption that both processes proceed over tens of milliseconds and are completed within about $100~\mathrm{ms}$, as a result of supersaturation depletion.
This is because accurate and precise estimation would require millisecond-scale measurements, which is challenging.
As a result. the estimated values for $A_n$ $k_g$ were $2.0\times10^{22}~\mathrm{m}^{-3}~\mathrm{s}^{-1}$ and $11.88~\mathrm{m}^{3}~\mathrm{kg}^{-1}~\mathrm{s}^{-1}$, respectively. The model satisfactorily captured the dynamic trend of both PSD and Z-average diameter (Figs.~\ref{fig:Fig. 4}ab). 
During the early stage, the Z-average diameter rapidly increased to approximately $120~\mathrm{nm}$ and remained stable without noticeable variation for tens of minutes.
After this period, particle size gradually increased to $200~\mathrm{nm}$, and resulting PSD was shifted to the larger region due to LNP agglomeration, which was captured by the coalescence model.

\subsection{Kinetic analysis}

\begin{figure}[H]
    \centering
    \includegraphics[width=0.9\textwidth]{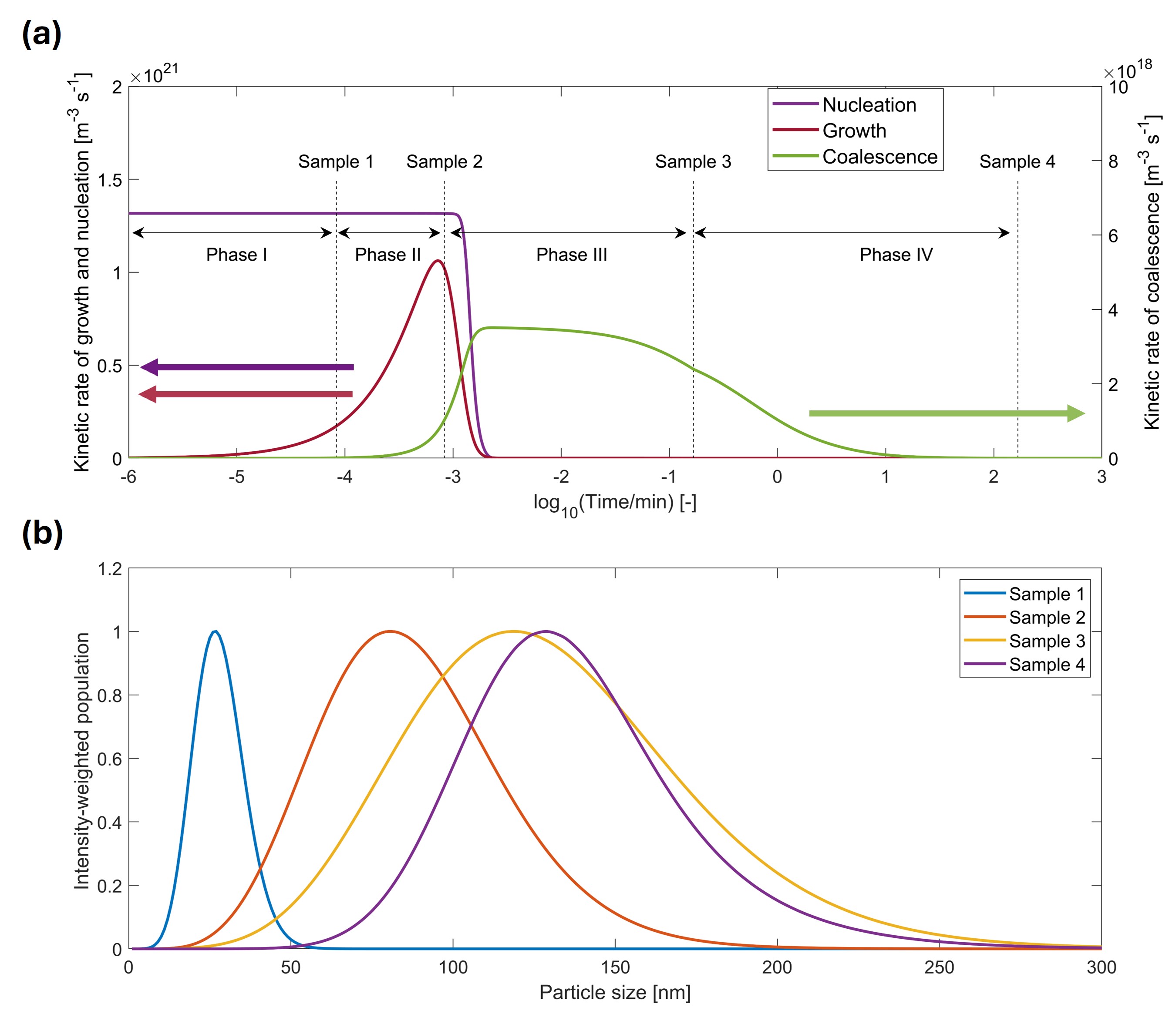}
    \caption{Contribution of individual kinetic rate during (a) LNP fabrication and (b) PSD sampled at the end of each phase.}
    \label{fig:Fig. 5}
\end{figure}

We investigated the contribution of individual kinetic processes to elucidate their effects on PSD evolution. To this end, the rate of each kinetic process was calculated and compared on a logarithmic scale throughout LNP fabrication (up to 1,000 min), as shown in Fig.~\ref{fig:Fig. 5}a. 
The absolute scale of nucleation and growth kinetic rate (left y-axis) was more than 100 times larger than coalescence kinetic rate (right y-axis), likely because PEG-lipid components served as barriers to particle collision, by causing steric hindrance.
The entire time frame was divided into four subphases based on its kinetic regime, and the PSD was sampled at the end of each subphase (Fig.~\ref{fig:Fig. 5}b).

In Phase I, nucleation was the dominant kinetic process, leading to a narrow PSD.
The gradual increase in growth rate during Phase II transformed the PSD into a log-normal-shaped distribution, likely due to the law of proportionate effect, as described in the Methods section.
In Phase III, both nucleation and growth ceased within 0.0015 min (about $100~\mathrm{ms}$) due to lipid depletion, and coalescence subsequently became the dominant process.
As shown in Fig.~\ref{fig:Fig. 3}b, the coalescence rate gradually declined thereafter, due to a decrease in attachment efficiency with increasing particle size.
As a result of the combined contribution of all three kinetic processes during Phase III, the PSD in Sample 3 became broader.
The broadened PSD observed in Sample 3 became narrower during Phase IV, where coalescence was the dominant kinetic process.
This trend can be attributed to the preferential coalescence of smaller particles, while larger particles tend to remain stable due to their higher energy barriers.
These findings suggest that each kinetic process plays a distinct role in shaping the PSD, highlighting the importance of systematically controlling each kinetic process for precise PSD tuning.
\subsection{Experimental observations of operation strategies for size control}

\begin{figure}[H]
    \centering
    \includegraphics[width=0.9\textwidth]{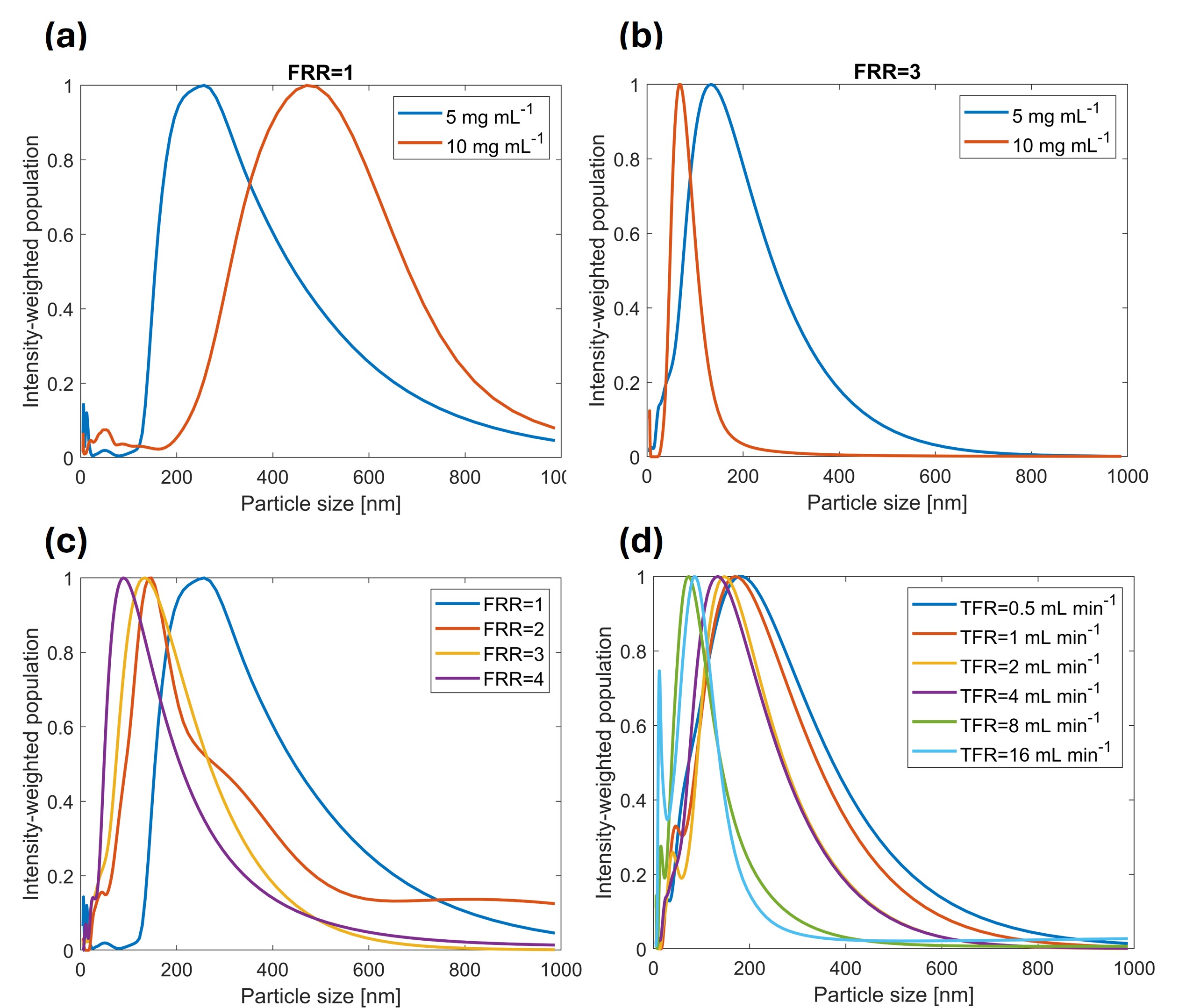}
    \caption{PSD measured at lipid concentrations of  $5$ and  $10 ~\mathrm{mg}~\mathrm{mL}^{-1}$ at (a) FRR = 1 and (b) FRR = 3. PSD measured at (c) FRR values from 1 to 4 and (d) TFR values from $0.5$ to $16~\mathrm{mL}~\mathrm{min}^{-1}$.}
    \label{fig:Fig. 6}
\end{figure}
In addition to examining the role of kinetic process in PSD evolution, this section presents a kinetic analysis of three commonly considered strategies for LNP size tuning: manipulation of (1) lipid concentration, (2) the flow rate ratio (FRR, defined as the aqueous-to-organic phase ratio), and (3) mixing rate. 
First, we experimentally examined the three strategies. The effect of mixing rate was analyzed by investigating the effect of total flow rate (TFR) because an increased TFR enhances mixing rate in the IJM. In this experiment, all PSDs were measured immediately after mixing in the IJM.

As lipid concentration in the organic phase was increased from $5$ to $10~\mathrm{mg}~\mathrm{mL}^{-1}$, two opposing effects were observed. For FRR = 1, the PSD shifted toward the larger region (Fig.~\ref{fig:Fig. 6}a), whereas it  shifted toward the smaller region for FRR = 3  (Fig.~\ref{fig:Fig. 6}b). When FRR was raised from 1 to 4 by increasing the aqueous phase volume, the PSD shifted to the smaller region (Fig.~\ref{fig:Fig. 6}c). Similarly, an increase in TFR from $0.5$ to $16~\mathrm{mL}~\mathrm{min}^{-1}$ also resulted in a PSD shift toward the smaller region (Fig.~\ref{fig:Fig. 6}d).

To further verify these observations, we compared our experiments with previously reported trends (Table~\ref{tab:Table. 1}). The size decreasing effects of both FRR and TFR were strongly consistent with the previously reported trend, whereas the effect of lipid concentration was inconsistent.
A size-increasing effect has been reported while we observed two opposing effects of lipid concentration. This can be attributed to the fact that solute concentration can have two opposing effects depending on the conditions in the nanoprecipitation system \cite{hamdallahMicrofluidicsPharmaceuticalNanoparticle2020}. Trends of the three operational strategies were frequently explained by the dilution of lipid and the resulting supersaturation level, which govern creation and growth of nuclei. These explanations align with conventional nanoprecipitation theory where the balance between nucleation and growth controls particle sizes \cite{hamdallahMicrofluidicsPharmaceuticalNanoparticle2020, nguyenControlledNucleationLipid2012}.
This suggests that the detailed mechanisms underlying these operational strategies can be interpreted in terms of kinetic processes under various mixing rate conditions.

\begin{landscape}
\begin{table}[h]
    \begin{center}
    \renewcommand{\arraystretch}{1.5}
    
     \begin{tabular}{|c|c|p{14cm}|}
        \hline
        \textbf{Operational strategy} & \textbf{Effect on particle size} & \textbf{Explanation} \\
        \hline
        \multirow{3}{*}{Increasing lipid concentration} 
            & Increase \cite{maekiUnderstandingFormationMechanism2017} & At higher lipid concentrations, lipid segments are more likely to encounter lipid molecules or other segments, leading to enhanced growth.  \\
        & Increase \cite{balbinoContinuousFlowProduction2013}  & Excessive amounts of lipid result in uncontrolled aggregation behavior \\
        & Increase \cite{Pradhan2008123123}&  \\
        \hline
        \multirow{7}{*}{Increasing FRR } 
            & Decrease \cite{maekiUnderstandingFormationMechanism2017} & Higher FRR has the same effect as low lipid concentration, resulting in smaller particle formation due to lipid dilution.  \\
        & Decrease \cite{okudaSizeregulationRNAloadedLipid2022} & Relative emissivity is raised by higher FRR, leading to agglomeration.  \\
        & Decrease \cite{zhigaltsevBottomUpDesignSynthesis2012}&  \\
        & Decrease \cite{maharjanComparativeStudyLipid2023}&  \\
        & Decrease \cite{rocesManufacturingConsiderationsDevelopment2020}&  \\
        & Decrease \cite{ripollOptimalSelfassemblyLipid2022}&  Relative emissivity affects energy landscape when FRR was increased. \\
        & Decrease \cite{choiPreciseControlLiposome2023}&  \\
        \hline
        \multirow{6}{*}{Increasing TFR} 
            & Decrease \cite{maekiMicrofluidicTechnologiesDevices2022} & Rapid dilution of lipid limits the fusion and growth of lipid segments. \\
        & No significant effect \cite{maharjanComparativeStudyLipid2023}& \\
        & Decrease \cite{ripollOptimalSelfassemblyLipid2022}&  \\
        & Decrease \cite{kimuraDevelopmentMicrofluidicBasedPostTreatment2020}&  \\
        & Decrease \cite{Pradhan2008123123}&  \\
        & Decrease \cite{choiPreciseControlLiposome2023}& Faster solvent depletion rate reduces particle size because disk-like lipid fragments have shorter time to grow. \\
        \hline
    \end{tabular}
    \end{center}
    
    \caption{Summary of the effects of key operational strategies on LNP size.}
    \label{tab:Table. 1}
\end{table}
\end{landscape}

\subsection{Kinetic interpretation of LNP size control}

Along with these experimental and theoretical evidence, we further focused on the kinetic processes, using the developed model. 

\subsubsection{Effect of lipid concentration}

\begin{figure}[H]
    \centering
    \includegraphics[width=0.9\textwidth]{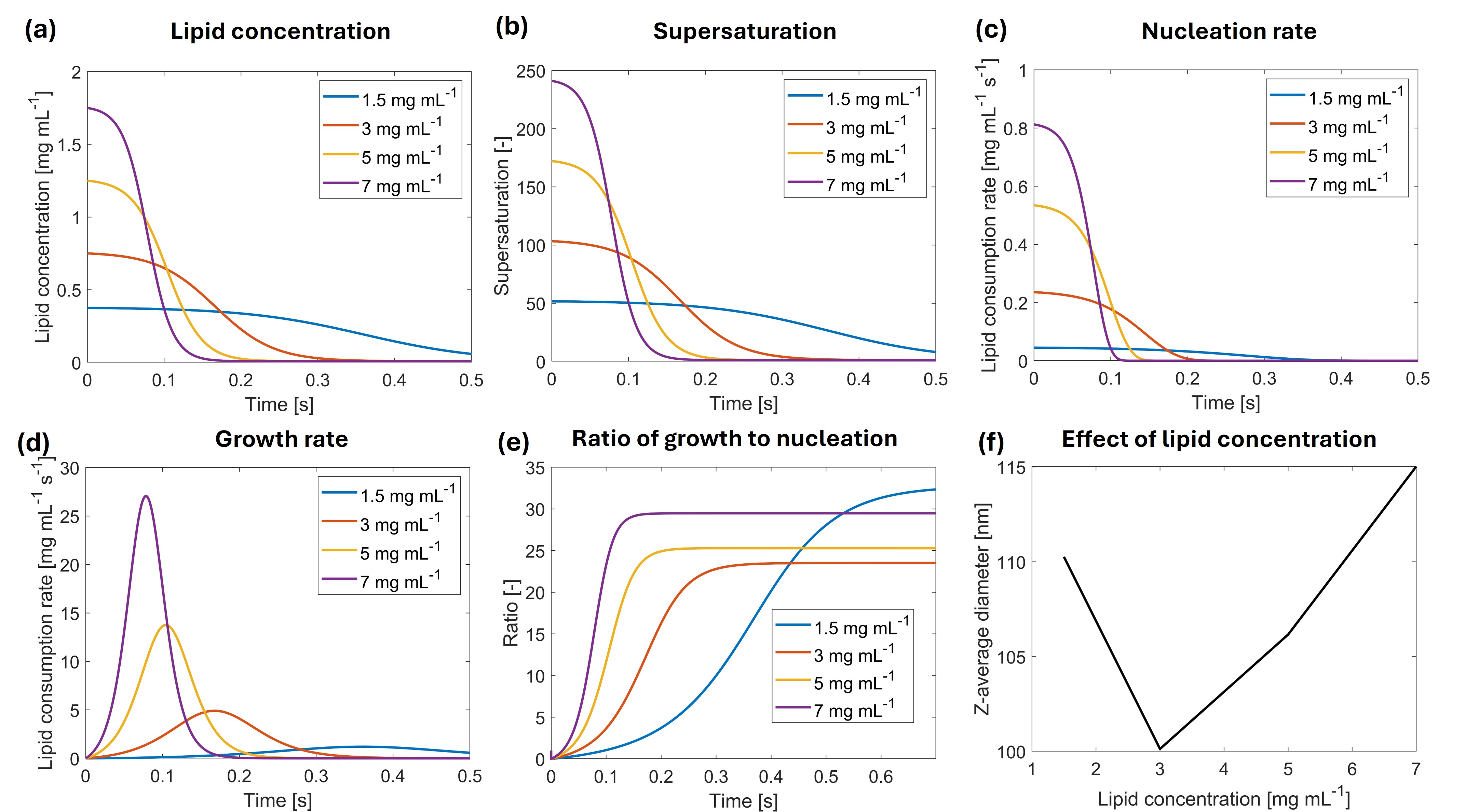}
    \caption{Dynamic profiles of (a) lipid concentration in the mixture, (b) supersaturation, lipid consumption rates for (c) nucleation and (d) growth, (e) ratio of lipid consumption for growth to nucleation, and (f) Z-average diameter  as lipid concentration increased from $1.5$ to $7~\mathrm{mg}~\mathrm{mL}^{-1}$.}
    \label{fig:Fig. 7}
\end{figure}

First, we investigate the effect of lipid concentration. The initial lipid concentrations in the ethanol solution are varied within the range of 1.5 to $7~\mathrm{mg}~\mathrm{mL}^{-1}$ for FRR = 3. Since the organic phase is assumed to be instantaneously mixed with the aqueous phase in our model, the lipid concentration is instantaneously diluted to 0.375, 0.475, 1.25, and $1.75~\mathrm{mg}~\mathrm{mL}^{-1}$, respectively (Fig.~\ref{fig:Fig. 7}a). 
As the lipid concentration is increased to $7~\mathrm{mg}~\mathrm{mL}^{-1}$, the initial supersaturation reaches 240 and rapidly declined, likely due to the depletion of lipid molecules (Fig.~\ref{fig:Fig. 7}b).
Accordingly, the lipid consumption rate for nucleation reaches its highest value of $0.8~\mathrm{mg}~\mathrm{mL}^{-1}~\mathrm{s}^{-1}$ (Fig.~\ref{fig:Fig. 7}c), suggesting an enhancement effect by lipid concentration. Similarly, the lipid consumption rate for growth is also significantly enhanced with a peak value of $27~\mathrm{mg}~\mathrm{mL}^{-1}~\mathrm{s}^{-1}$ (Fig.~\ref{fig:Fig. 7}d), confirming both nucleation and growth are significantly promoted by lipid concentration. 
However, the resulting particle size is determined by the relative balance between nucleation and growth kinetic rate. Therefore, the ratio of lipid consumption for growth to nucleation is investigated in Fig.~\ref{fig:Fig. 7}e to confirm how the balance affects LNP size. The ratio exhibits two opposing trends as the lipid concentration was increased. In the $1.5$ to $3~\mathrm{mg}~\mathrm{mL}^{-1}$ range of the lipid concentration, the converged ratio value decreases from 33 to 23, whereas it increases from 23 to 29 in the $3$ to $7~\mathrm{mg}~\mathrm{mL}^{-1}$ range of the lipid concentration. A similar trend is observed in the final Z-average diameter.
In the $1.5$ to $3~\mathrm{mg}~\mathrm{mL}^{-1}$ range, the particle size decreases from $110$  to $100~\mathrm{nm}$ (Fig.~\ref{fig:Fig. 7}f), suggesting that the increased supersaturation and nucleation rate outweighed growth. 
In contrast, in the $3$ to $7~\mathrm{mg}~\mathrm{mL}^{-1}$ range, the size increases from $100$ to $115~\mathrm{nm}$ (Fig.~\ref{fig:Fig. 7}f), implying that the promotion of growth exceeds that of nucleation, resulting in a convex particle size profile, which is a common trend in nanoprecipitation \cite{hamdallahMicrofluidicsPharmaceuticalNanoparticle2020}.
This multi-faceted role of lipid concentration may explain the two opposing effects observed in our experiments (Figs.~\ref{fig:Fig. 6}ab).

These multi-faceted effects imply that highly precise control is required to efficiently utilize lipid concentration as an operational parameter for LNP size tuning. Furthermore, decreasing the lipid concentration directly reduces the LNP production rate. In addition, a higher proportion of organic solvent increases the effort and time required for its removal per unit LNP product during post-processing. These findings highlight that manipulating lipid concentration might not be a universally effective strategy for LNP size control. Therefore, alternative operational strategies that effectively control particle sizes without compromising productivity should be  considered.

\subsubsection{Effect of FRR}

\begin{figure}[H]
    \centering
    \includegraphics[width=0.9\textwidth]{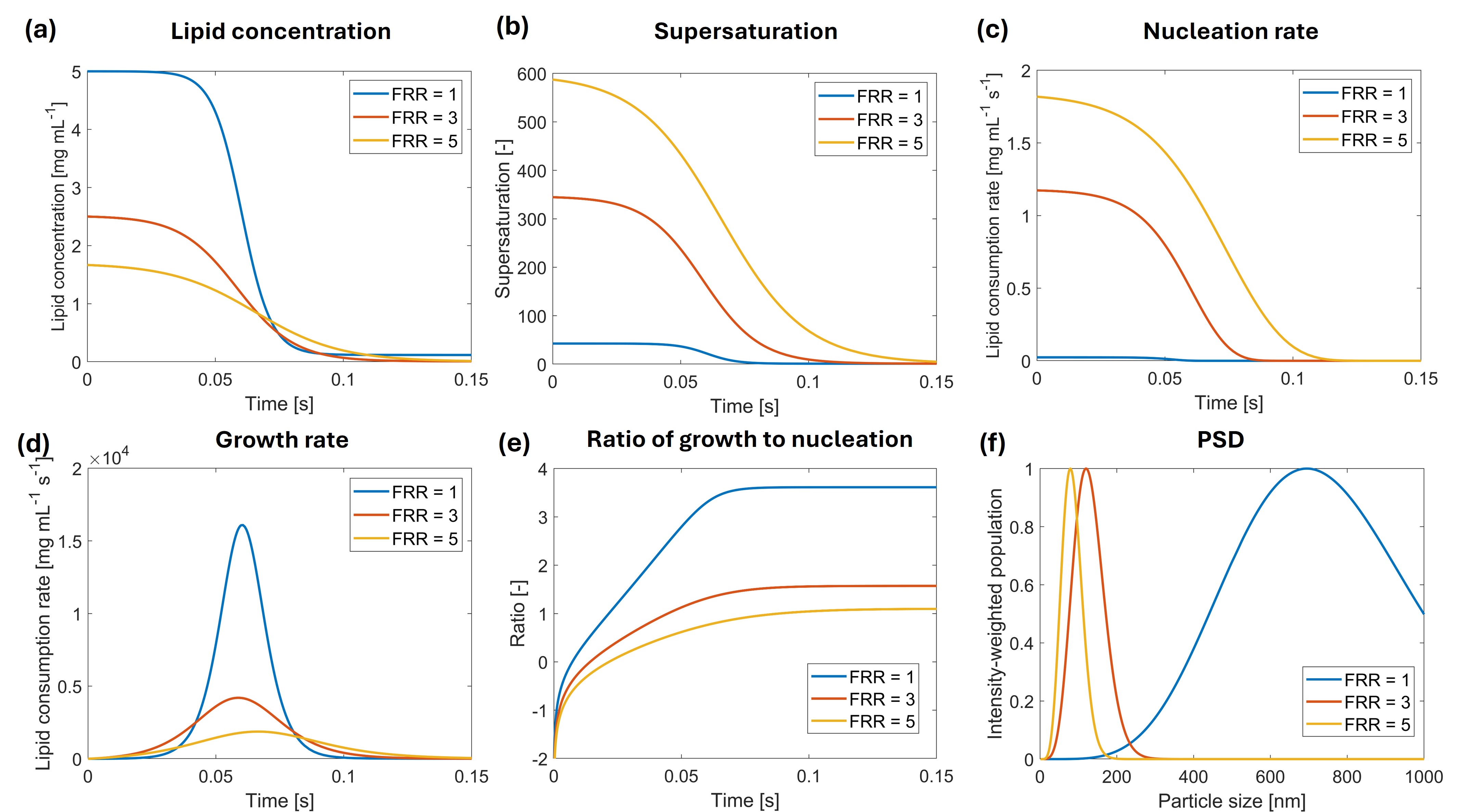}
    \caption{Dynamic profiles of (a) lipid concentration in the mixture, (b) supersaturation, lipid consumption rates for (c) nucleation and (d) growth, (e) ratio of lipid consumption for growth to nucleation, and (f) PSD for FRR = 1, 3, and 5.}
    \label{fig:Fig. 8}
\end{figure}

The effect of FRR is investigated by increasing its value from 1 to 5. The resulting initial lipid concentration at a given FRR is
\begin{equation}
    C_\text{lipid} = \frac{C_\text{lipid}^{0}}{\mathrm{FRR}+1}\text{,}
\end{equation}
where $C_\text{lipid}^{0}$ is the initial lipid concentration in the organic phase before mixed ($10~\mathrm{mg}~\mathrm{mL}^{-1}$).

Increasing the FRR to 5 significantly dilutes the initial lipid concentration to $1.6 ~\mathrm{mg}~\mathrm{mL}^{-1}$ (Fig.~\ref{fig:Fig. 8}a). However, the resulting supersaturation is elevated to 600 due to reduced lipid solubility in the mixture (Fig.~\ref{fig:Fig. 8}b). The elevated supersaturation level lasts for a longer duration of $0.15~\mathrm{s}$ than lower FRR. Consequently, the nucleation period is extended to 0.11 s (Fig.~\ref{fig:Fig. 8}c), likely leading to a greater number of new nuclei \cite{choiPreciseControlLiposome2023}.
Meanwhile, the lipid consumption rate for growth exhibits a smoother increase with a peak at $1.9 \times 10^{3}~\mathrm{mg}~\mathrm{mL}^{-1}~\mathrm{s}^{-1}$, which is only $12\%$ of that at FRR = 1 (Fig.~\ref{fig:Fig. 8}d). This weaker growth is probably attributed to the diluted lipid concentration and competition with enhanced nucleation. The enhanced nucleation and reduced growth rate are further confirmed by investigating the ratio of lipid consumption for growth to nucleation (Fig.~\ref{fig:Fig. 8}e).  
When the FRR is increased from 1 to 5, the ratio declines from 4,102 to 12, indicating that the kinetic balance is remarkably shifted toward nucleation. As a result, PSD shifts to the smaller region (Fig.~\ref{fig:Fig. 8}f) with Z-average diameter of $84~\mathrm{nm}$, which is an $89\%$ decrease compared to $752~\mathrm{nm}$ under FRR = 1.

These results demonstrate that the strategy of increasing FRR has two advantages in producing smaller particles: 
(1) An increased FRR elevates supersaturation due to low solubility, enhancing nucleation and promoting the formation of new nuclei.
(2) Under a higher FRR, lipid components are diluted to restrict particle growth without the need to introduce additional organic phase, which must be removed during post-processing.

\begin{figure}[H]
    \centering
    \includegraphics[width=0.9\textwidth]{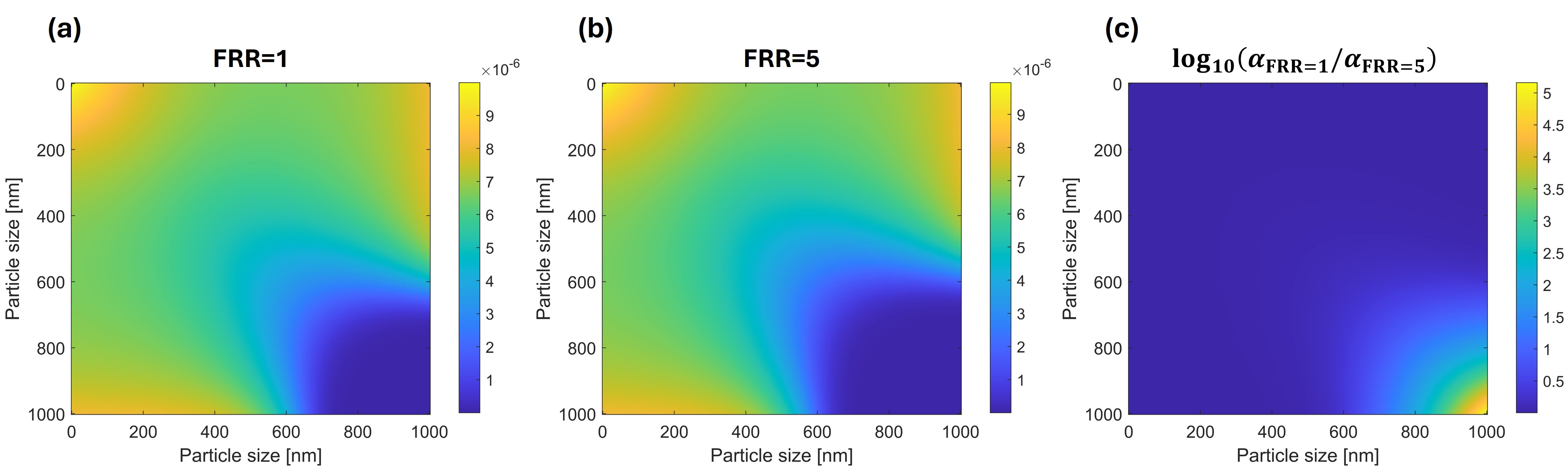}
    \caption{Attachment efficiencies under (a) FRR = 1 and (b) FRR = 5, respectively, and (c) shows the ratio of (a) to (b) on a logarithmic scale.}
    \label{fig:Fig. 9}
\end{figure}

We also investigate the effect of the mixed solvent's relative emissivity using DLVO theory, since higher FRR increases relative emissivity, thereby raising the energy barrier for LNP agglomeration \cite{ripollOptimalSelfassemblyLipid2022, okudaSizeregulationRNAloadedLipid2022}.
Specifically, we compare attachment efficiencies ($\alpha$), which control effective collisions based on energy barriers.
The mean relative emissivity is calculated as a function of FRR, 
\begin{equation}
    \epsilon_\text{mix} = \frac{\mathrm{FRR}}{\mathrm{FRR}+1}\epsilon_\text{water}+\frac{1}{\mathrm{FRR}+1}\epsilon_\text{ethanol}\text{,}
\end{equation}
where $\epsilon$ denotes relative emissivity, and the subscript mix, water, and ethanol correspond to each phase, respectively. The relative emissivity of water and ethanol are 78.2 and 24.3, respectively.
Based on this formulation, $\epsilon_\text{mix}$ is calculated to be 51 and 69 under FRR = 1 and 5, respectively. The attachment efficiencies were compared under FRR = 1 and FRR = 5 in Fig.~\ref{fig:Fig. 9}.

The ratio of $\alpha$ under FRR = 1 to that under FRR = 5 on a logarithmic scale is positive across all size conditions, indicating that $\alpha$ is higher under FRR = 1 than under FRR = 5 (Figs.~\ref{fig:Fig. 9}a--c).
The ratio significantly increases at larger particle size conditions (Fig.~\ref{fig:Fig. 9}c), suggesting that the effect of the mean relative emissivity would be more pronounced after large particles formed. Furthermore, the coalescence rate was relatively slow compared to nucleation and growth, as confirmed in the previous section. Therefore, the relative emissivity effect may have a significant impact in the incubation step over minute-to-hour timescales, rather than rapid self-assembly regime in the IJM step.

\begin{figure}[H]
    \centering
    \includegraphics[width=0.9\textwidth]{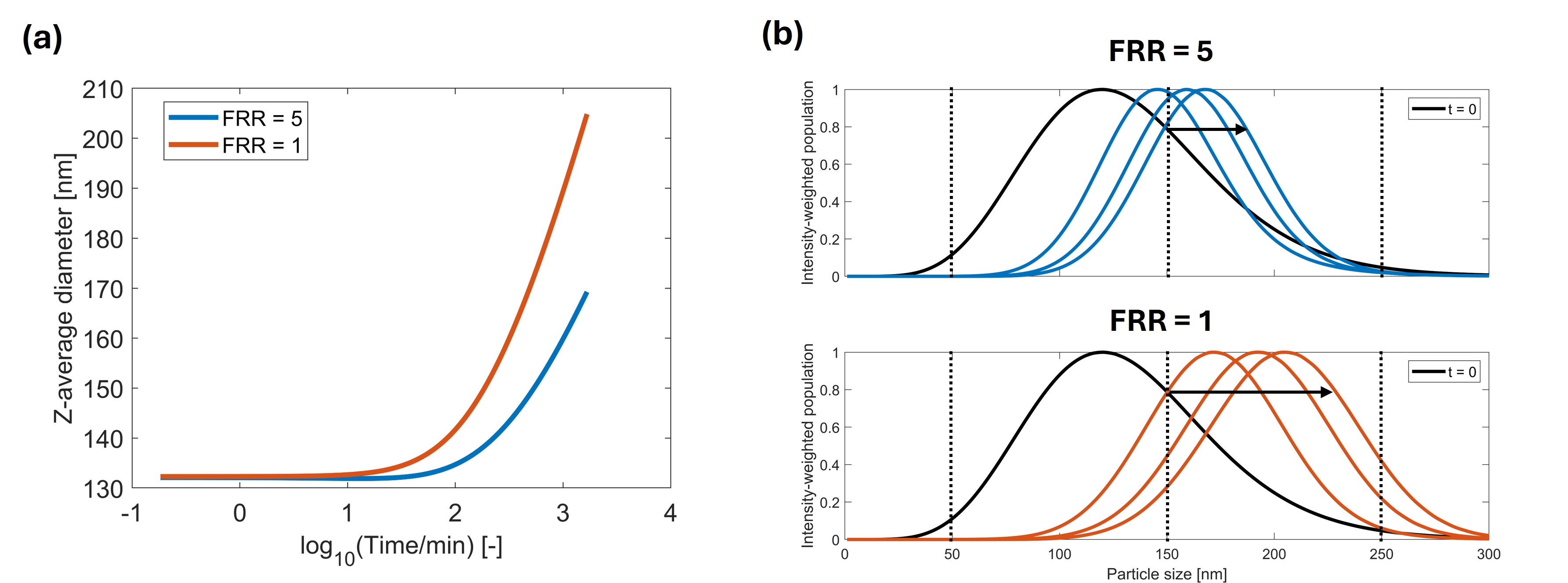}
    \caption{Dynamic profile of (a) the Z-average diameter at FRR = 5 and FRR = 1 and (b) PSDs investigated at time $t = 500$, 1000, and 1500 min for FRR = 5 and 1.}
    \label{fig:Fig. 10}
\end{figure}

To further investigate the long-term effect of FRR, we examine PBEs over 1,500 min ($25~\mathrm{h}$) under the coalescence kinetic regime. The dynamic profiles of Z-average diameter and PSDs are investigated under FRR = 1 and 5 with the same initial conditions. Under FRR = 1, the Z-average diameter reaches $204~\mathrm{nm}$, whereas it gradually increases to $166~\mathrm{nm}$ under FRR = 5, representing a $19\%$ reduction (Fig.~\ref{fig:Fig. 10}a). The same trend is observed in PSD profiles, showing that the shift of PSD is much faster under FRR = 1 than under FRR = 5 (Fig.~\ref{fig:Fig. 10}b). This suggests that the mean relative emissivity effect could be utilized for PSD tuning in post-processing steps. For example, the solvent proportion can be dynamically manipulated to optimize PSD during dialysis step, which takes approximately a day.

\subsubsection{Effect of mixing rate}

\begin{figure}[H]
    \centering
    \includegraphics[width=0.9\textwidth]{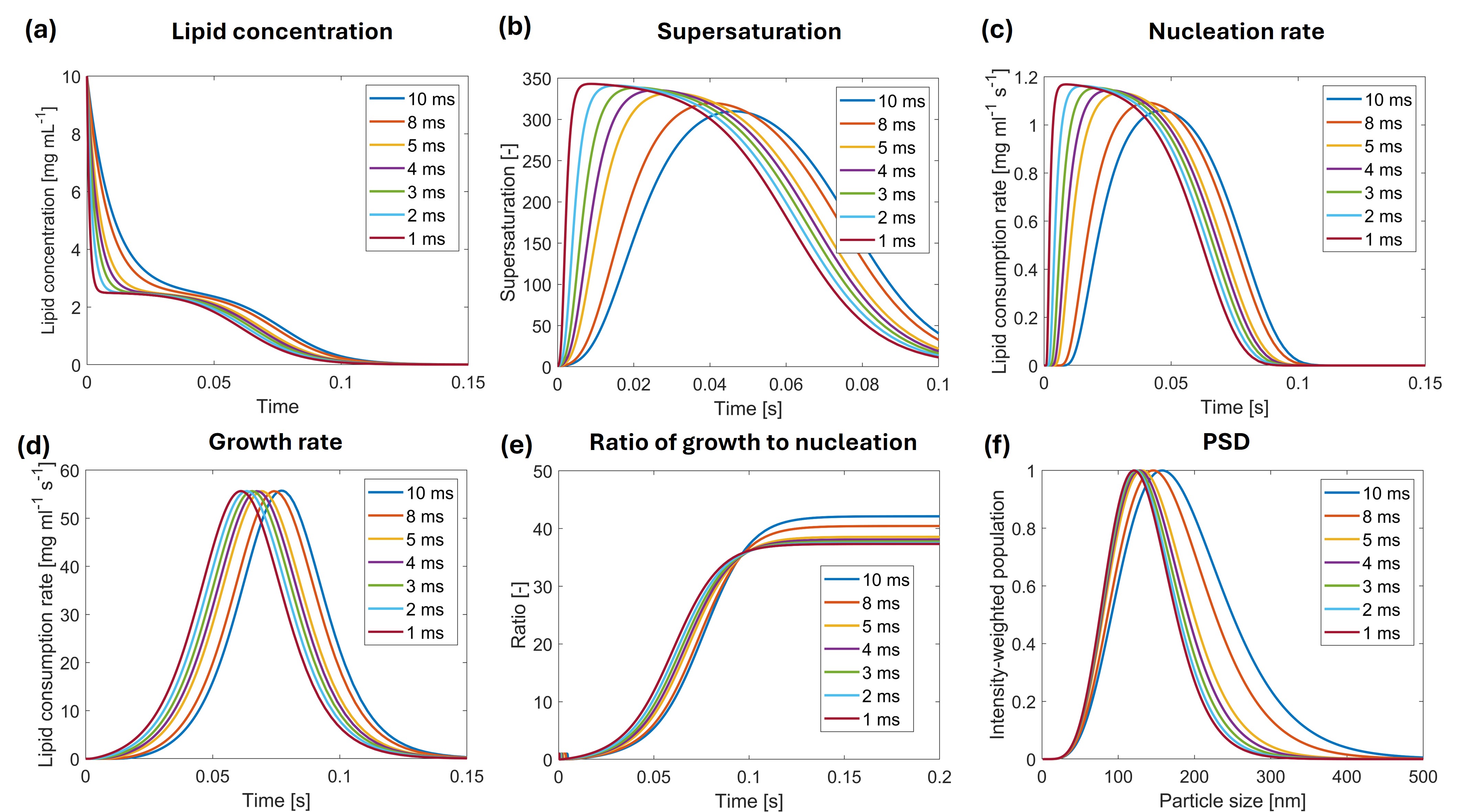}
    \caption{Dynamic profiles of lipid concentration in the mixture (a), supersaturation (b), lipid consumption rates for nucleation (c) and growth (d), ratio of lipid consumption for growth to nucleation (e), and PSD (f) as $\tau$ was decreased from $10~\mathrm{ms}$ to $1~\mathrm{ms}$.}
    \label{fig:Fig. 11}
\end{figure}

This section investigates how delayed mixing influences kinetic regime and increases LNP size. To numerically reflect delay in mixing, an exponentially converging function is introduced, which depends on the characteristic time scale of mixing ($\tau$). 
The volumetric fraction of aqueous buffer phase resulting from delayed mixing is given by
\begin{equation}
    f_\text{aqueous} = \frac{\mathrm{FRR}}{\mathrm{FRR}+1}\big(1-e^{-t/\tau}\big)\text{.}
\end{equation}
To increase the mixing rate, $\tau$ is decreased from $10$ to $1~\mathrm{ms}$ because smaller $\tau$ condition indicates faster mixing regime. The lipid concentration in the organic phase exhibits a two-stage decrease: the initial drop is due to delayed dilution with aqueous buffer, while the second decline is due to lipid consumption by nucleation and growth kinetic processes (Fig.~\ref{fig:Fig. 11}a).

At the slowest mixing condition of $\tau=10~\mathrm{ms}$,
the dilution process is significantly delayed, maintaining a high lipid concentration environment for an extended period. This delayed dilution results in a delayed supersaturation profile with a smoother curve and a peak value of 312 (Fig.~\ref{fig:Fig. 11}b).
As mixing is enhanced with $\tau = 1~\mathrm{ms}$, the lipid concentration almost promptly decreased within $0.05~\mathrm{s}$ due to rapid mixing. Accordingly, the supersaturation rapidly increases to 340 (Fig.~\ref{fig:Fig. 11}b). Consequently, the lipid consumption rate for nucleation is also promoted, peaking at $1.17~\mathrm{mg}~\mathrm{mL}^{-1}~\mathrm{s}^{-1}$ (Fig.~\ref{fig:Fig. 11}c), suggesting that more lipids are consumed in creating new nuclei. At this fastest mixing rate of $\tau = 1~\mathrm{ms}$, lipid consumption rate for growth shows its peak value at the earliest time of $0.059~\mathrm{s}$, whereas the peaking is delayed until $0.078~\mathrm{s}$ at the slowest mixing rate of $\tau = 10~\mathrm{ms}$ (Fig.~\ref{fig:Fig. 11}d). To investigate relative strength of nucleation and growth, the ratio of lipid consumption for growth to nucleation is calculated (Fig.~\ref{fig:Fig. 11}e). In the range until $0.09~\mathrm{s}$, faster mixing rate conditions show higher ratios. However, the trend is reversed after $0.09~\mathrm{s}$, suggesting that faster mixing rate results in a smaller particle size.   
As a result of the increase in mixing rate, the PSD is shifted to the smaller region (Fig.~\ref{fig:Fig. 11}f), and the Z-average diameter decreases from $186$ to $133~\mathrm{nm}$, representing a $28\%$ decrease. This confirms that enhanced mixing significantly decreases particle size.

Controlling the mixing rate is an effective strategy because it instantaneously achieves high supersaturation, thereby promoting nucleation rate. Unlike strategies involving lipid concentration or FRR, it does not require an additional liquid phase volume. This minimizes the effort and time required for solvent exchange during post-processing. Increasing the TFR not only enhances mixing rate but also improves production rate per unit volume of the system. These advantages suggest that the strategy of manipulating TFR is particularly effective for producing small particles.

These results confirm that the key to producing  small LNPs lies in (1) achieving high supersaturation and (2) controlling lipid dilution while minimizing additional organic solvent or aqueous buffer volume. Therefore, operational strategies such as channel geometry, injection method, or alternative organic solvents may be explored as alternative strategies to effectively manipulate supersaturation and lipid dilution.
\begin{figure}[H]
    \centering
    \includegraphics[width=0.5\textwidth]{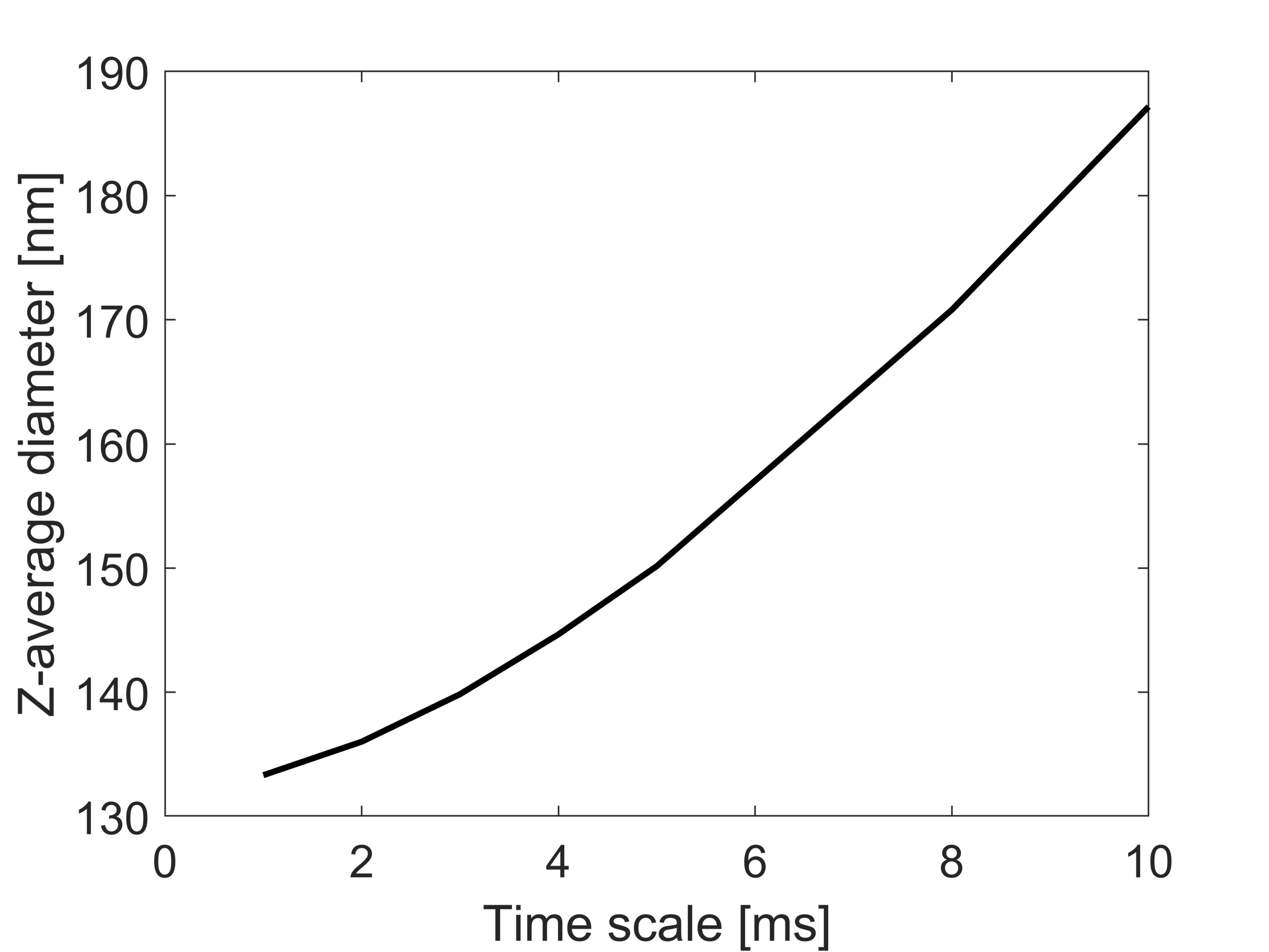}
    \caption{Summary plot of Z-average diameter as $\tau$ was increased from $1~\mathrm{ms}$ to $10~\mathrm{ms}$.}
    \label{fig:Fig. 12}
\end{figure}
To further illustrate the effect of mixing on particle size, we plot the Z-average diameter as a function of the characteristic mixing time scale ($\tau$) (Fig.~\ref{fig:Fig. 12}). Interestingly, a similar trend is observed in experiments \cite{choiPreciseControlLiposome2023}. 
In this study, the absolute scale of nucleation and growth kinetic parameters is roughly estimated from the PSD measured approximately at $1~\mathrm{min}$ after mixing, due to experimental limitations at the millisecond scale. However, the incorporation of $\tau$ as an input parameter may provide a direct way to 
simulate mixing effect on these kinetic processes. Since $\tau$ is a controllable variable and the corresponding particle sizes can be measured in experiment, we suggest that this approach enables indirect extraction of millisecond-scale information. Specifically, delayed mixing slows LNP formation, and stronger nucleation and growth would further amplify this effect, resulting in a steeper curve. Comparison between the model predictions and experiments will enable more precise kinetic parameter estimation.

\section{Conclusions}
This study presents a mechanistic model for understanding kinetic processes during LNP fabrication through nanoprecipitation. To predict PSD of LNPs, PBEs are formulated based on the nucleation, growth, and coalescence kinetic processes. The model, with the estimated kinetic parameters, satisfactorily captures the experimental trends in both PSD and Z-average diameter.

The individual kinetic rate is investigated to identify the role of each kinetic process on PSD shape. The scale of nucleation and growth kinetic rates is more than 100-fold larger than coalescence kinetic rate, which can be attributed to the stabilization effect posed by PEG-lipids.
In the early stage of LNP formation (up to $100~\mathrm{ms}$), the nucleation kinetic produces a narrow PSD, while growth kinetics transformed it to a log-normal-like shape. The coalescence kinetic gradually narrows the PSD in the long term.  

Based on these findings, the three most commonly considered operational strategies—manipulating lipid concentration, FRR, and mixing rate—are investigated both in experiments and the modeling study.

In the experiments, as the lipid concentration is increased from $5$ to $10~\mathrm{mg}~\mathrm{mL}^{-1}$, the PSD shifts to the larger region under $\mathrm{FRR}=1$, whereas the PSD shifts to the smaller region under $\mathrm{FRR}=3$, showing two opposing effects. When FRR is raised from 1 to 4 by increasing the
aqueous phase volume, the PSD shifts to the smaller region. Similarly, an increase in TFR from $0.5$
to $16~\mathrm{mL}~\mathrm{min^{-1}}$ also results in a PSD shift toward the smaller region.

The developed model successfully captures these trends and is used to elucidate detailed kinetic mechanisms underlying the three strategies. In the modeling study, the strategy of manipulating lipid concentration exhibits multi-faceted effects. When the lipid concentration is increased from $1.5$ to $7~\mathrm{mg}~\mathrm{mL}^{-1}$, both nucleation and growth rates are promoted, but the impact on the balance between nucleation and growth vary depending on the condition. In the $1.5$ to $3~\mathrm{mg}~\mathrm{mL}^{-1}$ range, the lipid consumption ratio for growth to nucleation exhibits a decreasing trend, consequently Z-average diameter also decreases from $110$ to $100~\mathrm{nm}$. On the other hand, the ratio exhibits an increasing trend, and Z diameter also increases from $100$ to $115~\mathrm{nm}$ in the $3$ to $7~\mathrm{mg}~\mathrm{mL}^{-1}$ range.
This suggests that lipid concentration should be carefully manipulated and that additional operational strategies should be simultaneously considered.

The strategy of increasing FRR produces significantly smaller particles through the increase in supersaturation. This promoted nucleation while suppressing growth, shifting the balance toward nucleation. As a result, when FRR is increased from 1 to 5, the Z-average diameter decreases from $752$ to $84~\mathrm{nm}$, which is an $89\%$ decrease.
Additionally, higher FRR restricted additional particle size increase during post-process after mixing, by increasing mean relative emissivity in the mixture. Specifically, increased emissivity raises electrostatic energy barrier, consequently decreasing attachment efficiencies and restricting coalescence rates. An increase in FRR from 1 to 5 suppresses particle size increase by $21\%$ over $25~\mathrm{h}$, indicating that the strategy of manipulating FRR can be utilized not only during the rapid self-assembly phase but also during the subsequent post-process.

The strategy of increasing mixing rate produces smaller particles, by rapidly raising supersaturation to promote formation of new nuclei. 
When the mixing is enhanced by decreasing characteristic time scale of mixing ($\tau$) from $10$ to $1~\mathrm{ms}$, the Z-average diameter decreases from $186$ to $133~\mathrm{nm}$, which is a $28\%$ decrease. Manipulation of mixing rate is an advantageous strategy in that it effectively controls supersaturation and nucleation rate without requiring additional solvent volume, which should be processed during dialysis.
We further propose the incorporation of $\tau$ via an exponential function, enabling the extraction of millisecond-scale information from experimental data.

We suggest that the key to producing small particles lies in controlling supersaturation and lipid dilution, while minimizing the volume of additional organic solvent or aqueous buffer.
Our findings provide mechanistic insights that are essential for developing optimization strategies for LNP size tuning.

\section*{CRedit authorship contribution statement}
\textbf{Sunkyu Shin:} Conceptualization, Methodology, Writing – original draft.
\textbf{Cedric Devos:} Investigation and Validation.
\textbf{Aniket Pradip Udepurkar:} Investigation and Validation.
\textbf{Pavan K. Inguva:} Conceptualization and Investigation.
\textbf{Allan S. Myerson:} Supervision.
\textbf{Richard D. Braatz:} Supervision.
\section*{Declaration of competing interest}
The authors declare that they have no known competing financial interests or personal relationships that could have appeared to influence the work reported in this paper.

\section*{Acknowledgments}
This research was supported by the U.S. Food and Drug Administration under the FDA BAA-22-00123 program, Award Number 75F40122C00200.

\bibliographystyle{elsarticle-num}
\bibliography{Elsevier}

\end{document}

\maketitle

\section{Implementation of the DLVO energy}
The van der Waals interactive energy ($\phi_{VDW}$) is calculated as follows:  
\begin{equation}
    \phi_{VDW} = -\frac{A_H}{6}\left(\frac{2r_1r_2}{a^2+2a(r_1+r_2)}+\frac{2r_1r_2}{a^2+2a(r_1+r_2)+4r_1r_2}+\ln{\frac{a^2+2a(r_1+r_2)}{a^2+2a(r_1+r_2)+4r_1r_2}}\right)
    \text{,}
\end{equation} where $A_H$, $r$, $a$ denote the Hamaker constant, particle radius, and interparticle distance, respectively.
In this study, the Hamaker constant is estimated to be $ 1.7\times10^{-22}~J$

The electrostatic repulsive energy ($\phi_{elec}$ is calculated as follows:
\begin{equation}
    \phi_{elec} = \frac{128\pi N_Ak_BT}{\kappa^2}\tanh^2{\left(\frac{e\psi}{4k_BT} \right)}\frac{r_{avg}}{r_{avg}+a}\exp({-\kappa a})
    \text{,}
\end{equation} 
where $N_A$, $k_B$, $\kappa$, $e$, $\psi$, $r_{avg}$ denote the Avogadro number, the Boltzmann constant, the Debye length, elementary charge, surface potential, average particle radius, respectively. 
In this study, the zeta potential value ($10~mV$) is used for the surface potential. 

The Debye length ($\kappa$) is calculated as follows:
\begin{equation}
    \kappa = \sqrt{\frac{2e^2N_AC}{e_0e_{rel}k_BT}}    \text{,}
\end{equation} 
where $C$, $e_0$, $e_{rel}$ denote ion concentration, emissivity in the vacuum, relative emissivity, respectively.